\documentclass[12pt,preprint]{aastex}
\newcommand \lsol{L$_{\odot}$}
\newcommand \msol{M$_{\odot}$}

\newcommand \mearth{M$_{\oplus}$}

\newfont{\rten}{cmr10}

\def \arcsec{\hbox{$^{\prime\prime}$}}

\begin{document}


\title{Are Debris Disks and Massive Planets Correlated?}

\author{Amaya Moro-Mart\'{\i}n\altaffilmark{1}, 
John M. Carpenter\altaffilmark{2}, 
Michael R. Meyer\altaffilmark{3}, 
Lynne A. Hillenbrand\altaffilmark{2}, 
Renu Malhotra\altaffilmark{4}, 
David Hollenbach\altaffilmark{5}, 
Joan Najita\altaffilmark{6}, 
Thomas Henning\altaffilmark{7}, 
Jinyoung S. Kim\altaffilmark{3}, 
Jeroen Bouwman\altaffilmark{7}, 
Murray D. Silverstone\altaffilmark{3} , 
Dean C. Hines\altaffilmark{8}, 
Sebastian Wolf\altaffilmark{7}, 
Illaria Pascucci\altaffilmark{3},
Eric E. Mamajek\altaffilmark{9} and
Jonathan Lunine\altaffilmark{4} 
}

\altaffiltext{1}{Department of Astrophysical Sciences, Peyton Hall, Ivy Lane, 
Princeton University, Princeton, NJ 08544, USA, amaya@astro.princeton.edu}

\altaffiltext{2}{Department of Astronomy, California Institute of Technology, 
Pasadena, CA 91125, USA}

\altaffiltext{3}{Steward Observatory, University of Arizona, 933 North Cherry Ave., 
Tucson, AZ 85721, USA}

\altaffiltext{4}{Department of Planetary Sciences, University of Arizona, 1629 E. 
University Boulevard, Tucson, AZ 85721, USA}

\altaffiltext{5}{NASA Ames, Moffet Field, CA 94035, USA}

\altaffiltext{6}{National Optical Astronomy Observatory, 950 North Cherry Ave.,  
Tucson AZ 85721, USA} 

\altaffiltext{7}{Max-Planck-Institut fur Astronomie, K\"onigstuhl 17, 69117 
Heidelberg, Germany} 

\altaffiltext{8}{Space Science Institute, 4750 Walnut Street, Suite 205, Boulder, CO 80301}

\altaffiltext{9}{Harvard-Smithsonian Center for Astrophysics, 60 Garden Street, MS-42, Cambridge, MA 02138}

\begin{abstract}
Using data from the $\it{Spitzer}$ $\it{Space}$ $\it{Telescope}$ 
Legacy Science Program ``Formation and Evolution of Planetary Systems'' (FEPS), we
have searched for debris disks around 9 FGK stars (2--10 Gyr), 
known from radial velocity (RV) studies to have one or more massive 
planets. Only one of the sources, HD 38529, has excess emission above the stellar photosphere; at 70 $\mu$m the 
signal-to-noise ratio in the excess is 4.7 while at $\lambda$ $<$ 30 $\mu$m 
there is no evidence of excess. The remaining sources show no excesses at any $\it{Spitzer}$ wavelengths. 
Applying survival tests to the FEPS sample and the results for the FGK survey published in 
Bryden et al. (\citeyear{bryd06}), we do not find a significant correlation between the frequency 
and properties of debris disks and the presence of close-in planets. We discuss possible 
reasons for the lack of a correlation. 

\end{abstract}


\section{Introduction}

During the last two decades, space-based infrared observations, first with $\it{IRAS}$ 
and then with $\it{ISO}$ and $\it{Spitzer}$, have shown that many main sequence stars are 
surrounded by dust disks (a.k.a. debris disks). These disks are generally observed by their 
infrared emission in excess over the stellar photosphere, but in some cases the disks have been 
spatially resolved and extend to 100s of AU from the central star. Dust particles are affected 
by radiation pressure, Poynting-Robertson and stellar wind drag, mutual collisions and collisions with 
interstellar grains, and all these processes contribute to make the lifetime of the dust 
particles significantly shorter than the age of the star/disk system. It is therefore thought 
that this dust is being replenished by a reservoir of undetected dust-producing planetesimals 
(Backman \& Paresce~\citeyear{back93}), like the asteroids, the Kuiper Belt Objects (KBOs) and 
the comets in our solar system. 
This represented a major leap in the search for other planetary 
systems: by 1983, a decade before extra-solar planets were discovered, $\it{IRAS}$ 
observations proved that there is planetary material surrounding nearby stars 
(Aumann et al.~\citeyear{auma84}). 

Preliminary results from $\it{Spitzer}$ observations of FGK (solar-type) stars 
indicate that the frequency of 24 $\mu$m excesses (tracing warm dust at asteroid 
belt-like distances) decreases from $\sim$30\%--40\% for ages $<$50 Myr,  
to $\sim$9\% for 100 Myr--200 Myr, and $\sim$1.2\% for ages $>$1 Gyr 
(Siegler et al.~\citeyear{sieg06}; Gorlova et al.~\citeyear{gorl06}; 
Stauffer et al.~\citeyear{stau05}; Beichman et al.~\citeyear{beic05}; 
Kim et al.~\citeyear{kim05}; Bryden et al.~\citeyear{bryd06}). 
On the other hand, Bryden et al. (\citeyear{bryd06}) estimated that the excess rate at 70 $\mu$m (tracing
colder dust at Kuiper Belt-like distances) is 13$\pm$5\% and is not correlated with stellar age on Gyr timescales. 
It is also found that FGK stars show large variations in the amount of excess emission at a given 
stellar age, and that the upper envelope of the ratio of the excess emission over the stellar photosphere 
at 24 $\mu$m decays as 1/$\it{t}$ for ages $>$20 Myr 
(Siegler et al.~\citeyear{sieg06}). 

These observations are consistent with numerical simulations of the evolution of dust generated 
from the collision of planetesimals around solar-type stars (Kenyon \& Bromley~\citeyear{keny05}).  
These models predict that after 1 Myr there is a steady 1/$\it{t}$ decline of the 24 $\mu$m~excess 
emission, as the dust-producing planetesimals get depleted. It is also found that this decay 
is punctuated by large spikes produced by individual collisional events between planetesimals
100--1000 km in size. These events initiate a collisional cascade leading to short-term increases 
in the density of small grains, which can increase the brightness density of the disk by an order of 
magnitude, in broad agreement with the high degree of debris disk variability observed by $\it{Spitzer}$
(Rieke et al.~\citeyear{riek05}; Siegler et al.~\citeyear{sieg06}). 

However, these models do not include the presence of massive planets, and the study of 
the evolution of the Solar System indicates that they may strongly affect 
the evolution of debris disk. There has been one major event in the early Solar System evolution 
that likely produced large quantities of dust. Between 4.5 Gyr and 3.85 Gyr ago there was a 
heavy cratering phase that resurfaced the Moon and the terrestrial planets, creating the lunar basins 
and leaving numerous impact craters on the Moon, Mercury and Mars. This ``Heavy Bombardment'' ended 
abruptly $\sim$3.85 Gyr ago, and since then the impact flux has been at least an order of magnitude 
smaller. During the last 20--200 Myr of the ``Heavy Bombardment'' epoch, a period known as 
the ``Late Heavy Bombardment'' (LHB), there was an increased cratering activity that came after 
a relatively calm period of several hundred million years, that could have been created by a sudden 
injection of impact objects from the main asteroid belt into the terrestrial zone. 
The orbits of these objects became unstable likely due to the the orbital migration of the giant 
planets which caused a resonance sweeping of the asteroid belt and a large scale ejection of 
asteroids into planet-crossing orbits (Strom et al.~\citeyear{stro05}; 
Gomes et al.~\citeyear{gome05}). This event, triggered by the migration of the giant planets,  
would have been accompanied by a high rate of asteroid collisions, and the corresponding high 
rate of dust production would have caused a large spike in the warm dust luminosity of the Solar System. 
Similarly, a massive clearing of planetesimals is thought to have occurred in the Kuiper Belt. 
This is inferred from estimates of the total mass in the KB region, 30--55 AU, ranging from
0.02~\mearth~(Bernstein et al.~\citeyear{bern04}) 
to $\sim$0.08~\mearth~(Luu \& Jewitt~\citeyear{luu02}), insufficient to have been able to 
form the KBOs within the age 
of the Solar System (Stern 1996). It is estimated that the primordial KB had a mass
of 30--50~\mearth~between 30-55 AU, and was heavily depleted after Neptune formed and
started to migrate outward (Malhotra, Duncan and Levison~\citeyear{malh00}; Levison et al.~\citeyear{levi06}). 
This resulted in the clearing of KBOs with perihelion distances near or inside the present orbit of 
Neptune, and in the excitation of the KBOs' orbits. The latter increased the relative velocities of 
KBOs from 10 m/s to $>$1km/s, making their collisions violent enough to result in a significant mass 
of the KBOs ground down to dust and blown away by radiation pressure. 

The evolution of debris disks may therefore be strongly 
affected by the presence of planets: it its early history, a star with planetary companions 
may be surrounded by a massive debris disk while the planets are undergoing orbital migration, 
whereas at a later stage the star would harbor a sparse dust disk 
after the dynamical rearrangement of the planets is complete (Meyer et al.~\citeyear{meye06a}); 
at very late stages, 2--10 Gyr, the production of dust may undergo occasional bursts due to major 
collisions of planetesimals stirred up by the planets. In addition to their effect on the dust production rates, massive 
planets can also affect the dynamics of the dust grains. Examples include the trapping of dust particles 
in mean motion resonances and their ejection due to gravitational scattering (Liou \& Zook~\citeyear{liou99}; 
Moro-Mart\'{\i}n \& Malhotra~\citeyear{ama02},~\citeyear{ama05}). 

In this paper, we search for debris disks around nine stars known 
from radial velocity studies to harbor one or more massive planets. These stars are drawn
from the $\it{Spitzer}$ Legacy program ``Formation and Evolution of Planetary Systems'' (FEPS).
The properties of the stars and their planetary companions can be found in Tables 1 and 2.
The observations and data reduction are briefly described in $\S$2 and the 
resulting spectral energy distributions (SEDs) are presented in $\S$3. In $\S$4 we explore the 
correlation of the frequency of dust emission with the presence of known planets by
applying survival tests to the FEPS
sample and the FGK-star survey published in Bryden et al. (\citeyear{bryd06}). 
Finally, $\S$5 discusses the interpretation of our results. 
HD38529, the only planet star in the FEPS sample with an excess emission,
is discussed in detail in Moro-Mart\'{\i}n et al. (in preparation). 

\section{Observations and Data Reduction}

An overview of the FEPS program is given in Meyer et al. (\citeyear{meye04}, 
\citeyear{meye06b}), and a detailed description of the data 
acquisition and data reduction in Hines et al. (\citeyear{hine05}) and 
Carpenter et al. (in preparation). The Multiband Imaging Photometer for 
$\it{Spitzer}$ (MIPS; Rieke et al.~\citeyear{riek04}) was used to obtain 
observations at 24 $\mu$m and 70 $\mu$m using the small field photometry
mode with 2-10 cycles of 3 and 10 s integration times, respectively. 
The data was first processed by the $\it{Spitzer}$ Science Center (SSC) 
pipeline version S13, and further processing was done by the FEPS team, the 
details of which can be found in Carpenter et al. (in preparation). 

At 24 $\mu$m, point-spread-function (PSF) fitting photometry was performed 
using the APEX module in MOPEX (Makovoz et al.~\citeyear{mako05}) using a fitting radius
of 21 pixels on the individual Basic Calibrated Data (BCD) images. Fluxes
were computed by integrating the PSF to a radius of 3 pixels, and then 
applying an aperture correction of 1.600 to place the photometry on the same
scale described in the MIPS Data Handbook.  The S13 images were processed 
using a calibration factor of 0.0447~MJy~sr$^{-1}$. We adopt a calibration 
uncertainty of 4\% as stated on the SSC MIPS web pages.

The raw MIPS~70\micron\ images were processed with the SSC pipeline
version S13. The individual
BCD images were formed into mosaics with 4\arcsec\ pixel sizes using the 
Germanium Reprocessing Tools (GeRT) software package S14.0 version 1.1 
developed by the SSC. The GeRT package performs column filtering on the BCD 
images to remove streaks in the BCD images, and then performs a time median 
filter to remove residual pixel response variations. A $40''\times40''$ region 
centered on the source position was masked when computing the time and column 
filtering such that the filtering process is not biased by the source. The 
filtered images were formed into mosaics using MOPEX. 
Aperture photometry was performed on the 
MIPS~70\micron\ mosaics using a custom modified version of IDLPHOT. The 
adopted aperture radius of 16\arcsec\ was chosen to optimize the signal to 
noise for faint sources. The sky-level was computed as the mean value of the 
pixels in a sky-annulus that extends from 40$''$ to 60$''$. The photometry 
uncertainty is given by $\sigma$ = $\Omega$$\sigma$$_{sky}$$\sqrt{\it{N_{ap}}}$$\eta$$_{sky}$$\eta$$_{corr}$
$\sqrt{1.0+\it{N_{ap}/N_{sky}}}$, 
where $\sigma$$_{sky}$ is the standard deviation in the sky annulus surface 
brightness,  $\Omega$ is the pixel solid angle, $\it{N_{sky}}$ and $\it{N_{ap}}$ are the number
of pixels in the sky annulus and in the aperture, and 
$\eta$$_{sky}$ and $\eta$$_{corr}$ are correction factors that account for the presence on the 
mosaic of non-uniform noise and of correlated noise, respectively. 
We used $\eta$$_{sky}$ = 2.5 and  $\eta$$_{corr}$ = 1.40
(see full description in Carperter et al. in preparation). 
The adopted calibration factor is 702~MJy sr$^{-1}$ / (DN s$^{-1}$) with an 
uncertainty of 7\% as described on the SSC MIPS web pages. 

The Infrared Spectrograph (IRS; Houck et al.~\citeyear{houc04}) was used to 
obtain low-resolution (R = 70--120) spectra from 7.4 $\mu$m to 38 $\mu$m, with 
integration times per exposure of 6 s and 14 s for the Short-Low 
(7.4--14.5 $\mu$m) and Long-Low (14.0--38.0 $\mu$m), respectively. 
The data was initially processed with the SSC pipeline S10.5.0, with further
processing described in Bouwnman et al. (in preparation). From the spectra, 
synthetic photometric points were calculated at 13 $\mu$m with a rectangular
bandpass between 12.4$\mu$m and 14.0$\mu$m, at 24$\mu$m with the same 
bandpass shape as the MIPS~24 filter, and at 33 $\mu$m with a rectangular
bandpass between 30 $\mu$m and 35 $\mu$m. 
The estimated calibration uncertainty in the synthetic 
photometric is 6\% (Carpenter et al. in preparation). The spectra 
are generally  not reliable beyond 35 $\mu$m, although we found that for 
HD 6434, HD 121504 and HD 80606 it is very noisy beyond 34 $\mu$m, 
33 $\mu$m and 30$\mu$m, respectively, making the 33 $\mu$m photometric 
points unreliable for the last two sources. The latter is flaged 
by the SSC as non-nominal, possibly due to a failure in the peak-up. 

The Infrared Array Camera (IRAC; Fazio et al.~\citeyear{fazi04}) was used to 
obtain observations at 3.6 $\mu$m, 4.5 $\mu$m and 8.0 $\mu$m in subarray mode.
Initial processing of the data was done with the SSC pipeline S13, with 
further processing as described in Carpenter et al. (in preparation). 
Aperture photometry on individual IRAC frames were performed using a custom 
modified version of IDLPHOT using an aperture radius of 3 pixels (1 pixel 
$\sim$ 1.2\arcsec), with the background annulus extending from 10 to 20 pixels
centered to the star. The internal uncertainty was estimated as the standard 
deviation of the mean of the photometry measured at the four dither positions.
We adopted calibration factors of 0.1088, 0.1388, and 0.2021 MJy/sr per DN/s 
for IRAC 3.6, 4.5\micron\ and 8\micron\ respectively and calibration 
uncertainties of 2\% (Reach et al.~\citeyear{reac05}).

\section{Spectral Energy Distributions and Excess Emission}

The SEDs are shown in Fig. 1 and include the $\it{Spitzer}$ photometric 
measurements, observations made by $\it{IRAS}$, Tycho and 2MASS
and, in some cases, upper limits at 1.2 mm from Carpenter et al. (\citeyear{carp05}). 
For each star, the results from the  $\it{Spitzer}$ photometric measurements and their internal 
uncertainties are listed in the first entry of Table 3 (in rows indicated by ``obs''). 
The reported fluxes arise from both the photosphere of the star and the thermal
emission of the dust (if present). For all targets, observations are sufficient 
to detect the photosphere of the star at all $\it{Spitzer}$ wavelengths $<$33 $\mu$m, making it
possible to detect small dust excesses (limited mainly by the calibration uncertainties). 
To estimate the contribution from the dust alone, we need to subtract the photospheric emission, 
given in the second entry of Table 3 (under ``model''). The Kurucz model calculations are 
described in Carpenter et al. (in preparation). Infrared excesses can also be identified 
empirically from the color-color diagrams in Fig. 2, showing a narrow distribution of the ratio of the 
fluxes at 24 $\mu$m and 8 $\mu$m (F$_{24}$/F$_{8}$) compared to a wide distribution 
of the ratio of the fluxes at 70 $\mu$m and 24 $\mu$m (F$_{70}$/F$_{24}$). This indicates that 
the flux at 24 $\mu$m is mainly photospheric and that the best indicator of the presence of a 
debris disks is the 70 $\mu$m excess emission. 

HD 38529 has the only robust detection of an excess at 70 $\mu$m with a signal-to-noise (SNR) in the excess
of 4.7, a small excess at 33 $\mu$m (with a SNR in the excess of 0.7) 
and no excess $<$30 $\mu$m. Because of the slope of the 
spectrum across the MIPS 70 $\mu$m~band, the 70 $\mu$m flux needs to be color-corrected by dividing 
the observed flux by 0.893 (assuming the emission arises from cold dust emitting like a blackbody 
at 50 K - see MIPS Data Handbook). This increases the 70 $\mu$m flux from 75.3 mJy to 84.3 mJy. 
The centroid positions of the object in the 24 $\mu$m and 70 $\mu$m images are RA=05:46:34.88, 
DEC=+01:10:04.61 and RA=05:46:34.79, DEC=+01:10:04.62, respectively, in 
agreement with the 2004.7 2MASS coordinates for HD 38529 (RA=05:46:34.895 and DEC=+01:10:04.65; 
accounting for the proper motion of the star), and with the absolute pointing knowledge:  
better than 1.4\arcsec~and 1.7\arcsec~(1-$\sigma$ radial) at 24 $\mu$m and 70 $\mu$m, 
respectively ($\it{Spitzer}$ Observers Manual). 
Inspection of the images show that the 24~$\mu$m~and 70~$\mu$m~source is free of
nearby point sources, and there is very little structure from galactic
cirrus. Finally, it is unlikely that the emission 
at 70 $\mu$m comes from a background galaxy within 2\arcsec of the stellar 
position: from the background counts in Dole et al.~\citeyear{dole04} and 
following Downes et al. (\citeyear{down86}) we estimate a probability of
1.5$\times$10$^{-5}$ for 50 mJy and 7.4$\times$10$^{-6}$ for 100 mJy. We therefore conclude that
the observed 70 $\mu$m emission comes from HD 38529. Even though it is difficult to
identify statistical trends  from one detection, it is interesting to note 
that HD 38529 is the most luminous, most massive, and most evolved of the planet
bearing stars in Table 1. Assuming V = 5.95 (Johnson 1966), a Hipparcos distance
of 42 pc and no reddening, the object has an absolute visual magnitude of M$_{v}$ = 
2.81 and Log(L/\lsol) = 0.82, putting the star on the Hertzsprung gap, so 
it is clearly post-main sequence. 

In summary, we find that only 1 out of 9 of the planet-bearing stars show evidence 
of a debris disk. In the next section, we explore whether or not there is evidence 
of a correlation between the presence of debris disks and close-in planets.

\section{Are Debris Disks and Close-in Planets Related Phenomena?}

Using the FEPS data, we address the possibility of a debris/planet connection 
by comparing the results for the nine planet-bearing stars (hereafter ``planet sample'')
to those of a larger sub-set of stars in the FEPS sample without regard to the presence
of planets (hereafter ``control sample''). The planet sample is a subset of the 
control sample. Given the current statistics from RV surveys, it is unlikely that
the majority of the stars in the control sample harbor a giant planet, therefore, 
the control sample is likely less biased for the presence of giant planets than is
the planet sample. By comparing these two samples, we investigate if the 
frequency and luminosity of debris disks are correlated with the presence of a 
massive planet. 

\subsection{Selection of the control sample}

The main criterion for the choice of the control sample is that the observations reached similar
levels of sensitivity as for the planet sample. For this we require that:
(1) the stars in the control sample span the same range of distances as the planet 
bearing stars (26--62 pc); and
(2) their infrared background levels at 70 $\mu$m are similar. 
Age may also be a factor, as the stars in the FEPS sample range from 3 Myr to 3 Gyr typically, 
with a few stars perhaps as old as 3--10 Gyr, while planet 
bearing stars are typically older than 1 Gyr. If debris disks evolve significantly over 
the range 3 Myr--10 Gyr, this could also introduce a bias in the debris disk detection. 
Samples of young stars show an initial rapid decline over the first $\sim$100 Myr, while 
Bryden et al. (\citeyear{bryd06}) found that the excess rate at 70 $\mu$m
is 13$\pm$5\% (down to a fractional luminosity of L$_{dust}$/L$_*$$\sim$10$^{-5}$, i.e. about 100 times 
the luminosity of the Kuiper Belt dust) and is not correlated with stellar age on Gyr timescales. 
Therefore, by choosing a control sample that is restricted to stars older than 300 Myr,  
we do not expect to introduce any significant age bias, while improving the statistics by 
increasing the number of stars in the control sample. 
Our control sample thus consists of 99 stars with distances 26--62 pc and 
ages $>$300 Myr. We use the Kolmogorov-Smirnov (K-S) test (see e.g. Press et al.~\citeyear{pres93}) to
assess whether or not the distributions of distances and IR background levels of the planet and the 
control samples are consistent with having been drawn from the same parent population. 
The K-S test yields two values: $\it{D}$, a measure of the largest difference between the 
two cumulative distributions under consideration and $\it{Probability(D>observed)}$, 
an estimate of the significance level of the observed value of $\it{D}$ as a disproof 
of the null hypothesis that the distributions come from the same parent population, 
i.e. a very small value of $\it{Probability(D>observed)}$ implies that the distributions
are significantly different. Because in this case we find $\it{Probability(D>observed)}$ = 0.6 
(for distance) and $\it{Probability(D>observed)}$ = 0.4 (for IR background), we conclude that 
both samples could have been drawn from the same distribution in terms of distance and IR background
levels and therefore can be compared. Note that the K-S test for age yield a much lower probability 
($\it{Probability(D>observed)}$ $\sim$10$^{-5}$), i.e. both samples are likely not drawn
from the same distribution in terms of age. However, given that the observations indicate
that for the ages under consideration ($>$ 300 Myr, with approximately half
of the stars having ages $>$ 1 Gyr) there is no correlation between the 
70 $\mu$m~excess and the stellar age, we do not expect to introduce any significant bias 
by comparing both samples (but keep in mind that the validity 
of the comparison relies on the observed lack of correlation with age).

\subsection{Frequency of Debris Disks}

With respect to the frequency of debris disks, we find that 1/9 stars in the FEPS planet
sample has 70 $\mu$m excess emission with a SNR in the excess $>$3, compared to 9/99 
stars in the FEPS control sample; for the Bryden et al. (\citeyear{bryd06}) survey, the 
rates are 1/11 (planet sample) and 7/69 (control sample). 
Because the frequency of debris disks (seen at 70 $\mu$m) in the planet sample and the control sample 
are similar, we conclude that there is no evidence of the presence of a correlation between the 
frequency of debris disks and close-in planets (if we were to assume a $\sqrt{\it{N}}$ 
error in the number of stars with excesses, the frequency of debris around
a planet-bearing star would be within a factor of 3 of the control sample). 
At 24 $\mu$m (tracing warmer dust), 
the frequency of debris disks could also be similar in the planet and the control samples, 
as none of the stars in the FEPS planet sample show excess emission, while 2/99 stars in the
FEPS control sample do. 

\subsection{Fractional Excess Luminosity: Survival Analysis}

The planet sample and the control sample are dominated by upper limits, therefore, 
the K-S test is not sufficient to assess the probability that they could
have been drawn from the same parent distribution. 
To extract the maximum amount of information from the non-detections
it is necessary to use survival analysis methods, which make certain
assumptions about the underlying distributions. 
Using ASURV Rev 1.2 (LaValley et al. ~\citeyear{lava92}), which implements the survival 
analysis methods of Feigelson \& Nelson (\citeyear{feig85}), we have used the Gehan, 
logrank and Peto-Prentice tests to compute the probability that the planet sample and the 
control sample could have been drawn from the same parent distribution with respect to 
the fractional excess luminosity, L$_{dust}$/L$_{\star}$. 

We use the fractional luminosity of the excess, L$_{dust}$/L$_{\star}$, instead of the 
70 $\mu$m excess flux to minimize any correlation with distance. Following Bryden 
et al. (\citeyear{bryd06}), from the 70 $\mu$m excess emission one can estimate 
the fractional luminosity of the excess by assuming a single dust temperature, 
T$_{dust}$ = 52.7 K, corresponding to an emission peak at 70 $\mu$m. In this
case, L$_{dust}$/L$_{\star}$ $\sim$ 10$^{-5}$(5600/T$_*$)$^3$(F$_{70,dust}$/F$_{70,*}$), 
where F$_{70,dust}$ and F$_{70,*}$ are the dust excess and photospheric flux at 
70~$\mu$m and T$_*$ is the 
stellar temperature in units of Kelvin . For non-detections, F$_{70,dust}$ = 3$\times$$\Delta$F$_{70}$, 
where $\Delta$F$_{70}$ is the 1-$\sigma$~uncertainty of the observed flux. 

The resulting survival analysis probabilities, using 3-$\sigma$ upper limits, 
are 0.64 (Gehan), 0.86 (logrank) and 0.72 (Peto-Prentice). 
As discussed in  Feigelson \& Nelson (\citeyear{feig85}), the logrank test is 
more sensitive to differences at low values of the variable under consideration
(i.e. near the upper limits), while the Gehan test is more sensitive to 
differences at the high end (i.e. for the detections). The Peto-Prentice test is 
preferred when the upper limits dominate and the sizes of the samples to be compared
differ (as it is our case). Similarly, we have carried out survival analysis for the sample of 69 FGK main sequence 
stars in Bryden et al. (\citeyear{bryd06}). This sample was selected with regard to expected 
signal-to-noise ratio for stellar photospheres and is not biased for or against known planet-bearing stars. 
The planet sample consists of 11 stars with known close-in planets and the control 
sample includes all 69 stars. With respect to the FEPS targets, these stars are generally closer and the 
observations are therefore sensitive to less luminous debris disks (see Fig. 3). 
In this case, the probabilities that the planet sample and the 
control sample could have been drawn from the same parent distribution with respect to 
the fractional excess luminosity are  0.83 (Gehan), 0.86 (logrank) and 0.70 (Peto-Prentice).
If we consider the FEPS and Bryden's samples together, these probabilities are
0.62, 0.85 and 0.70, respectively. 
Because all the probabilities are larger than 0.6, i.e. significantly larger than 0,
the conclusion from the Gehan, logrank and Peto-Prentice tests 
from the data collected so far (from both FEPS and the GTO results in Bryden et al.~\citeyear{bryd06}), 
is that we cannot rule out the hypothesis that the planet sample and
the control sample have been drawn from the same population with respect to the fractional excess 
luminosity. In other words, we find no sign of correlation between the excess luminosity 
and the presence of close-in massive planets. 

\section{Discussion}

\subsection{Comparison to Previous Studies}

Greaves et al. (\citeyear{grea04}) searched for submillimeter dust emission around 8 stars
known from radial velocity studies to have giant planets orbiting within a few AU, and 
found no debris disks down to a dust mass limit of 
6$\times$10$^{-8}$~\msol\footnote{Dust mass estimates for the KB dust disk range from a total dust 
mass $<$ 3$\times$10$^{-10}$~\msol~(Backman et al.~\citeyear{back05}) to 
$\sim$ 4$\times$10$^{-11}$~\msol~for dust particles $<$ 150 $\mu$m 
(Moro-Mart\'{\i}n \& Malhotra~\citeyear{ama03}); with a 
fractional luminosity of L$_{dust}$/L$_*$$\sim$10$^{-7}$--10$^{-6}$ (Stern~\citeyear{ster96}). 
The fractional luminosity of the asteroid belt dust (a.k.a zodiacal cloud) is estimated to 
be L$_{dust}$/L$_*$$\sim$10$^{-8}$--10$^{-7}$ (Dermott et al.~\citeyear{derm02}).}; they also noted that out of 
20 solar-type stars known to have disks, only one, $\epsilon$ Eridani, 
has a planet orbiting inside a few AU (Hatzes et al.~\citeyear{hatz00}), 
concluding that either debris disks and close-in giant planets are unrelated phenomena 
or they are mutually exclusive. However, these results had severe limitations due to 
the low sensitivity of the submillimeter observations and the limited accuracy of the 
radial velocity measurements due to the youth of the sample.
The high sensitivity of the mid-IR observations with $\it{Spitzer}$ allow a fresh 
reexamination of the question.
Preliminary work from a $\it{Spitzer}$/MIPS GTO program by Beichman et al. (\citeyear{beic05})
found that out of 26 FGK field stars known to have planets through radial velocity studies, 
six stars (HD 33636, HD 50554, HD 52265, HD 82943, HD 117176 and HD 128311)  
show 70~$\mu$m excess with a SNR in the excess of 12.4, 12.1, 3.2 15.7, 3.2 and 4.6, respectively, 
implying the presence of cool dust ($<$100 K) located mainly beyond 10 AU. 
These six stars have a median age of 4 Gyr and their fractional luminosities, 
L$_{dust}$/L$_{star}$, in the range (0.1--1.2)$\times$10$^{-4}$
are $\sim$100 times that inferred for the KB (Beichman et al.~\citeyear{beic05}). 
The study suggested a tentative correlation between the frequency and magnitude of 
the dust emission with the presence of known planets. Our analysis of the FEPS and 
the Bryden et al. (\citeyear{bryd06}) samples do not confirm the presence of such a correlation. 

\subsection{Interpretation of our Results}

We found that, given the $\it{Spitzer}$ and radial velocity data we have so far, 
there is no evidence of a correlation between the presence of close-in massive planets
and the frequency and excess luminosity of debris disks, i.e. debris disks are not
more prevalent in systems with close-in massive planets than in systems selected without 
regard to whether they have close-in massive planets or not. 

This might seem a surprising result because it is reasonable to assume that most giant 
planets formed in systems that were initially rich in planetesimals, as plantesimals are 
the building blocks of giant planets in the core accretion model. 
However, despite a likely initial abundance of dust-producing planetesimals, systems 
with giant planets may not produce abundant debris at Gyr ages. 
The solar system is one example where there is significant evidence that
it once had a massive planetesimal belt despite the little debris produced today. 
That is, giant planets may  play an important role in the evolution of debris disks 
by efficiently grinding away or ejecting planetesimals from an initially massive disk. 
This could involve processes similar to the LHB in the early Solar System, where a large 
fraction of the dust producing planetesimals were lost due to the orbital migration of the giant planets. 
For the planet-bearing stars, given that the conditions for the formation of at least one giant planet 
were met, we could speculate that additional massive planets possibly formed and migrated, which could 
lead to LHB-type of events. Comparison of the debris disk properties between stars
with and without massive giant planets may therefore be a function of age. The FEPS and 
Bryden et al. (\citeyear{bryd06}) samples are drawn mainly from stars 300 Myr to 10 Gyr old, 
i.e. mostly after the LHB is thought to have occurred in our Solar System. 

Our result also suggests that massive planets may not be required to produced debris. 
A possible mechanism for the production of debris in the presence or 
absence of massive planets is the collisional cascade model of Kenyon \& Bromley (\citeyear{keny05}). 
Such a model can produce debris at Gyr ages, even in disks that are too low in solids to form 
a giant planet (i.e. too low in initial disk mass and /or too low in metallicity). In this model,
large 1000 km size planetesimals can stir up smaller planetesimals (0.1--10 km in size) along 
their orbits, starting a collisional cascade that can produce dust excess emission of 
the magnitude shown in Fig. 3 over the relevant range of ages. 
However, this cannot be the only mechanism because if it were to dominate debris production
one would expect to see the dust temperature to be correlated with age and this
trend has not been observed (Najita \& Williams~\citeyear{naji05}). Similarly, the 
observations in Fig. 3 could not confirm the time dependence of the fractional 
70 $\mu$m excess emission predicted by the models. 

That massive planets may not be required to produce debris is also supported by several observational 
results.
Firstly, debris disks are more common than massive planets: 
it is found that $>$7\% of stars have giant planets with M$<$13 M$_{Jup}$ and semimajor axis
within 5 AU, but this is a lower limit because the duration of the surveys (6--8 years) 
limits the ability to detect planets between 3 AU and 5 AU. The expected frequency of gas giant planets increases 
to $\sim$12\% when RV surveys are extrapolated to 20 AU (Marcy et al.~\citeyear{marc05}), with the
distribution of planets following d$\it{N}$/d$\it{M}$ $\propto$$\it{M}$$^{-1.05}$ from M$_{Saturn}$ to 
10 M$_{Jup}$ (the surveys are incomplete at smaller masses).  
In comparison, 
the frequency of debris disks observed at 70 $\mu$m with $\it{Spitzer}$ is 
13$\pm$5\% (from Bryden et al.~\citeyear{bryd06}). 
However, this detection rate is sensitivity limited because the observations in Bryden et al. (\citeyear{bryd06})
can only reach fractional luminosities of L$_{dust}$/L$_*$$\gtrsim$10$^{-5}$, i.e. 
$\gtrsim$100 times the luminosity from our Solar System Kuiper Belt.
Bryden et al. (\citeyear{bryd06}) found that the frequency of dust detection increases steeply as 
smaller fractional luminosities are considered, going from nearly 0\% for L$_{dust}$/L$_*$$\sim$10$^{-3}$, 
to 2$\pm$2\% for L$_{dust}$/L$_*$$\sim$10$^{-4}$ and 13$\pm$5\% for L$_{dust}$/L$_*$$\sim$10$^{-5}$.  
Using this cumulative distribution and assuming that the distribution of debris disk 
luminosities is a Gaussian, Bryden et al. (\citeyear{bryd06}) estimated that the luminosity of the Solar System dust
is consistent with being 10 $\times$ brighter or fainter than an average solar-type star, i.e. debris disks 
at the Solar System level could be common. The debris disks observed with $\it{Spitzer}$ could therefore be
the high luminosity tail of a distribution of dust luminosities that peaks near the Solar System values. 

Secondly, there is no correlation between stellar metallicities and the incidence of debris disks
(Beichman et al.~\citeyear{beic05}, Bryden et al.~\citeyear{bryd06} and
Greaves, Fischer \& Wyatt~\citeyear{grea06}). 
Greaves, Fischer \& Wyatt (\citeyear{grea06}) found that in a
sample of 310 F7--K3 stars within 25 pc of the Sun and for which the stellar 
metallicities are known, there is only a 0.6\% probability that planet-bearing
stars and debris disks stars have the same metallicity distribution, with the planet-bearing
stars being correlated with high stellar metallicities (Fischer \& Valenti~\citeyear{fisc05}).
This is in agreement with the core accretion model, where the formation of giant planets
requires the presence of a large surface density of solids in the disk, so that the
planet can grow a core sufficiently large to accrete an atmosphere before the 
gas disk disappears in $\lesssim$10 Myr. Because the governing time scale in the growth 
of planetesimals is the orbital period, in the KB region the planetesimal formation 
process is slower (according to Kenyon and Bromley~\citeyear{keny04} it may take 
$\sim$ 3 Gyr to form a Pluto at 100 AU), but can proceed well after the 
gas disk has dissipated (so there is no time limitation). This can occur in systems
regardless of whether or not they meet the conditions for giant planet formation. 

As a result, collisional grinding in a self-stirred model of this kind might 
be expected to produce debris in systems with low metallicities 
and low initial disk masses.  If this leads to debris production in a 
wider variety of systems than can produce giant planets, we might 
expect the presence of debris to be poorly correlated with the presence 
of giant planets.

\section{Conclusions}

\begin{itemize}
\item Using $\it{Spitzer}$ observations, we have searched for debris disks 
around 9 planet-bearing solar-type stars, with stellar ages ranging from 2 to 10 Gyr. 
Only one of the sources, HD 38529, is found to have excess emission above the 
stellar photosphere, with a signal-to-noise ratio at 70 $\mu$m of 4.7 and 
no excess at $\lambda$ $<$ 30 $\mu$m. The remaining sources show no excesses at any 
$\it{Spitzer}$ wavelengths.  
\item Given the data we have so far, from both FEPS and the FGK sample from Bryden et al. 
(\citeyear{bryd06}), and using survival analysis, we find that there is no evidence 
of a correlation between the presence of close-in massive planets and the
frequency and excess luminosity of debris disks. 
\item Because we expect massive planets to form in systems that are initially 
rich in planetesimals, but the observations indicate that systems with giant planets do 
not preferentially show debris, there is the possibility that massive planets play an important 
role in the evolution of debris disks by efficiently grinding away or 
ejecting planetesimals from an initially massive disk, possibly in a LHB-type of event. 
\item Our result also suggests that massive planets may not be required to produced debris, which
is supported by the collisional cascade models of Kenyon \& Bromley (\citeyear{keny05}), and
the observations and theoretical models that indicate that debris disks are more 
prevalent than massive planets. 
\end{itemize}


\begin{center}
{\bf APPENDIX: HD 150706}
\end{center}

HD 150706 is part of the FEPS sample and exhibits an excess emission at 
70 $\mu$m with a SNR in the excess of 4.3 (and a color-corrected flux of 46.3 mJy). 
Even though it has been listed as a planet-bearing star, it is not included in our planet sample 
because new radial velocity observations cannot confirm the 
claimed planet. HD~150706 has appeared in various compilations of sun-like stars with 
extra-solar planets (c.f. Santos, Israelian, Mayor \citeyear{sant04}). 
An orbital solution for a purported 1.0 $M_{Jup}$ eccentric planet 
at 0.8 AU was announced by the Geneva Extrasolar Planet Search Team 
(2002, Washington conference ``Scientific Frontiers in Research 
in Extrasolar Planets"; Udry, Mayor \& Queloz, \citeyear{udry03}); 
however, there is no refereed discovery paper 
giving details, only web pages. Eight doppler velocity measurements (Table 4)
made with HIRES on the Keck telescope from 2002 to 2006 yield RMS of 12.1 m/s, 
far below the 33 m/s claimed velocity amplitude due to a planet.  The RMS for a linear fit 
of the HIRES data is 8 m/s which can be adequately explained 
by the expected jitter for a young (700$\pm$300 Myr) and active early G star like HD~150706.  
The four years of HIRES data rule out the presence 
of planets of roughly 1 M$_{Jup}$ or larger located within 2 AU, and 
2 M$_{Jup}$ or more within 5 AU (modulo sin($\it{i}$)). Smaller planets 
inward of 5 AU, and super-Jupiters outward of 5 AU are not inconsistent 
with the HIRES observations to date. Further, the lack of 
a monotonic trend in the velocities of amplitude many tens of m/s indicates 
that there is no brown dwarf nor a low mass star anywhere within $\sim$ 20 AU. 

\begin{center} {\it Acknowledgments} \end{center}
We thank the rest of the FEPS team members, colleagues at the $\it{Spitzer}$ Science Center 
(in particular Dave Frayer), and members of all the $\it{Spitzer}$ instrument teams for advice 
and support. We thank G. Marcy, R.P. Butler, S.S.Vogt, and D.A. Fischer for their work with Keck/HIRES
and the anonymous referee for useful comments. 
This work is based on observations made with the $\it{Spitzer}$ Space Telescope,
which is operated by the Jet Propulsion Laboratory, California Institute of Technology 
under NASA contrast 1407. 
A.M.M. is under contract with the Jet Propulsion Laboratory (JPL) funded by NASA through
the Michelson Fellowship Program. A.M.M. is also supported by the Lyman Spitzer Fellowship at Princeton University.
M.R.M. and R.M. are supported in part through the LAPLACE node of NASA's Astrobiology Institute.
R.M. also acknowledges support from NASA-Origins of Solar Systems research program.
S.W. was supported through the DFG Emmy Noether grant WO 875/2-1 and WO875/2-2.
FEPS is pleased to acknowledge support from NASA contracts 1224768 and 1224566
administered through JPL.

\clearpage

\begin{deluxetable}{lllllllll}
\tablewidth{0pc}
\tablecaption{Stellar properties}
\tablehead{
\colhead{Source} & 
\colhead{Spectral} &
\colhead{Distance\tablenotemark{a}} &
\colhead{Age} &
\colhead{T$_{eff}$} &
\colhead{log(L)} &
\colhead{M} &
\colhead{[Fe/H]}\\
\colhead{(HD $\#$)} & 
\colhead{Type} &
\colhead{(pc)} &
\colhead{(Gyr)} &
\colhead{(K)} &
\colhead{(log(L$_{\odot}$))} &
\colhead{(M$_{\odot}$)} & 
\colhead{}}

\startdata
 6434    & G2/3V\tablenotemark{d} & 40$\pm$1 & 12$\pm$1\tablenotemark{e} & 5835\tablenotemark{g} & 0.05$\pm$0.02\tablenotemark{c} & 0.84$\pm$0.05\tablenotemark{f} &  -0.52\tablenotemark{g}\\  

 38529   & G8III/IV\tablenotemark{d}  & 42$\pm$2 & 3.5$\pm$1\tablenotemark{h} & 5697\tablenotemark{i} & 0.80\tablenotemark{i} & 1.47\tablenotemark{i}&  0.445\tablenotemark{i}\\  

 80606   & G5\tablenotemark{j}  & 58$\pm$20 & 6\tablenotemark{k} & 5573\tablenotemark{i} & -0.15\tablenotemark{i} & 1.06\tablenotemark{i}&  0.343\tablenotemark{i}\\  

 92788   & G6V\tablenotemark{l}  & 32$\pm$1 & 6$\pm$2\tablenotemark{m} & 5836\tablenotemark{i} & 0.01\tablenotemark{i} & 1.13\tablenotemark{i}&  0.318\tablenotemark{i}\\  

 106252  & G0\tablenotemark{j}  & 37$\pm$1 & 5.5$\pm$1\tablenotemark{n} & 5870\tablenotemark{i} & 0.11\tablenotemark{i} & 1.01\tablenotemark{i}&  -0.076\tablenotemark{i}\\  

 121504  & G2V\tablenotemark{o}  & 44$\pm$2 & 2$\pm$1\tablenotemark{p} & 6075\tablenotemark{g} & 0.19$\pm$0.04\tablenotemark{c} & 1.03$\pm$0.06\tablenotemark{f}&0.16\tablenotemark{g} \\  

 141937  & G2/3V\tablenotemark{q}   & 33$\pm$1 & 2.6$\pm$1\tablenotemark{r} & 5847\tablenotemark{i} & 0.07\tablenotemark{i} & 1.08\tablenotemark{i}&  0.129\tablenotemark{i}\\  

 179949  & F8V\tablenotemark{q}   & 27$\pm$1 & 2.5$\pm$1\tablenotemark{s} & 6168\tablenotemark{i} & 0.27\tablenotemark{i} & 1.21\tablenotemark{i}& 0.137\tablenotemark{i}\\  
 
 190228  & G5IV\tablenotemark{t}   & 62$\pm$3 & 5\tablenotemark{s} & 5348\tablenotemark{i} & 0.63\tablenotemark{i} & 1.21\tablenotemark{i}& -0.180\tablenotemark{i} \\  

\tablenotetext{~}{References: 
(a) HIPPARCOS Catalogue, Perryman et al. (1997);
(c) Computed from FEPS database;
(d) Houk (1980); 
(e) Barbieri \& Gratton (2002),  Nordstrom et al. (2004), Ibukiyama \& Arimoto (2002); 
(f) Nordstrom et al. (2004); 
(g) Santos et al. (2004); 
(h) Valenti \& Fischer (2005), Gonzalez et al. (2001);
(i) Valenti \& Fischer (2005); 
(j) Cannon \& Pickering (1918-24); 
(k) Mamajek (in preparation);
(l) Houk \& Smoth-Moore (1999); 
(m) Wright et al. (2004); Laws et al. (2003); Gonzalez et al. (2001);
(n) Valenti \& Fischer (2005), Wright et al. (2004), Laws et al. (2003), Mamajek (in preparation);
(o) Houk \& Cowley (1975); 
(p) Barbieri \& Gratton (2002), Mamajek (in preparation);
(q) Houk \& Smoth-Moore (1988); 
(r)  Barbieri \& Gratton (2002), Nordstrom et al. (2004), Laws et al. (2003), Valenti \& Fischer (2005), Mamajek (in preparation);
(s)  Nordstrom et al. (2004), Valenti \& Fischer (2005);
(t) Jaschek (1978)}
\enddata
\end{deluxetable}

\clearpage

\begin{deluxetable}{lllllll}
\tablewidth{0pc}
\tablecaption{Orbital characteristics of known planetary companions}
\tablehead{
\colhead{Planet} & 
\colhead{M$_{p}$ sin$\it{i}$} &
\colhead{Period} &
\colhead{a$_{p}$} &
\colhead{e$_{p}$} &
\colhead{N$_{obs}$} &
\colhead{Ref.}\\
\colhead{(HD $\#$)} & 
\colhead{(M$_{Jup}$)} &
\colhead{(days)} &
\colhead{(AU)} &
\colhead{} &
\colhead{} &
\colhead{}} 
\startdata
 6434b     & 0.397(59)       & 21.9980(90)   & 0.1421(82) & 0.170(30)   & 130 & (1)\\  

 38529b    & 0.852(74)       & 14.3093(13)   & 0.1313(76) & 0.248(23)   & 162 & (2)\\  
 38529c    & 13.2(1.1)       & 2165(14)      & 3.72(22)   & 0.3506(85)  & 162 & (2)\\  

 80606b    & 4.31(35)        & 111.4487(32)  & 0.468(27)  & 0.9349(23)  & 46  & (2)\\
 \nodata   & 3.90(9)         & 111.81(23)    & 0.47       & 0.9227(12)  & 61  & (3)\\

 92788b    & 3.67(30)        & 325.81(26)    & 0.965(56)  & 0.334(11)   & 58  & (2)\\
 \nodata   & 3.58 	       & 325.0(5)      & 0.96       & 0.35(1)     & 55  & (1)\\

 106252b   & 7.10(65)        & 1516(26)      & 2.60(15)   & 0.586(65)   & 15  & (2)\\
 \nodata   & 7.56            & 1600(18)      & 2.7        & 0.471(28)   & 40  & (4)\\

 121504b   & 1.22(17)        & 63.330(30)    & 0.329(19)  & 0.030(10)   & 100 & (1)\\

 141937b   & 9.8(1.4)        & 653.2(1.2)    & 1.525(88)  & 0.410(10)   & 81  & (5)\\

 179949b   & 0.916(76)       & 3.092514(32)  & 0.0443(26) & 0.022(15)   & 88  & (2)\\

 190228b   & 4.49	       & 1146(16)      & 2.25       & 0.499(30)   & 51  & (4)\\

\tablenotetext{~}{HD 38529 has 2 known planets. The multiple
entries for the other stars correspond to different estimates for the same
planet. $\it{a}$$_{p}$ and $\it{e}$$_{p}$ are the semimajor axis and eccentricity of the 
planet. N$_{obs}$ is the number of radial velocity observations. The number in parenthesis indicate the 
uncertainty in the last significant figures.}
\tablenotetext{~}{References:
(1) Mayor et al. (2004);
(2) Butler et al. (2006); 
(3) Naef et al. (2001);
(4) Perrier et al. (2003);
(5) Udry et al. (2002).}
\enddata
\end{deluxetable}

\clearpage

\begin{deluxetable}{llllllllllll}
\rotate
\tablewidth{0pc}
\tablecaption{$\it{Spitzer}$ photometry and Kurucz stellar models for FEPS targets with planets}
\tablehead{
\colhead{Source} & 
\colhead{} & 
\colhead{IRAC 3.6} & 
\colhead{IRAC 4.5} & 
\colhead{IRAC 8} & 
\colhead{IRS 13} & 
\colhead{MIPS 24} & 
\colhead{IRS 24} & 
\colhead{IRS 33} & 
\colhead{MIPS 70} & 
\colhead{SNR$_{exc}$} & 
\colhead{1200}\\
\colhead{(HD $\#$)} &
\colhead{} & 
\colhead{($\mu$m)} & 
\colhead{($\mu$m)} & 
\colhead{($\mu$m)} & 
\colhead{($\mu$m)} & 
\colhead{($\mu$m)} & 
\colhead{($\mu$m)} & 
\colhead{($\mu$m)} & 
\colhead{($\mu$m)} & 
\colhead{at 70$\mu$m} & 
\colhead{($\mu$m)}}
\startdata
6434     & obs          & 952$\pm$7      & 603$\pm$7   & 215$\pm$1     & 74.9$\pm$0.8 & 23.9$\pm$0.2 & 25.0$\pm$0.8  & 11$\pm$1 	     & 8.0$\pm$7.4 & 0.7              & 0$\pm$10 \\ 


& model & 892$\pm$28     & 565$\pm$17  & 203$\pm$6     & 73$\pm$2   & 23.2$\pm$0.7 & 23.2$\pm$0.7   & 12.2$\pm$0.4       & 2.6$\pm$0.1  & & \nodata \\


 38529 & obs          & 5893$\pm$42     & 3634$\pm$44 & 1340$\pm$9    & 467$\pm$5  & 150$\pm$1& 146$\pm$2  & 86$\pm$2       & 75$\pm$11 & 4.7 & \nodata\\

& model                 & 5935$\pm$283   & 3689$\pm$168& 1360$\pm$66   & 487$\pm$24 & 156$\pm$8    & 156$\pm$8      & 82$\pm$4           & 17.4$\pm$0.8 & & \nodata\\


 80606 & obs           & 340$\pm$2      & 210$\pm$3   & 75.5$\pm$0.5  & 23.8$\pm$0.7 & 8.65$\pm$0.08  & 7.9$\pm$0.3  &  3.1$\pm$0.6       & 3$\pm$5 & 0.5      & \nodata\\

& model                 & 335$\pm$12     & 208$\pm$7   & 77$\pm$3      & 27$\pm$1   & 8.8$\pm$0.3  & 8.8$\pm$0.3    & 4.6$\pm$0.2        & 0.98$\pm$0.04& & \nodata \\     


 92788 & obs           & 1447$\pm$10    & 891$\pm$11  & 323$\pm$2     & 111$\pm$1  & 36.1$\pm$0.3 & 37$\pm$1       & 20$\pm$2           & 11$\pm$9 & 0.8    & 5$\pm$15\\


& model                 & 1438$\pm$52    & 902$\pm$31  & 329$\pm$12    & 118$\pm$4  & 38$\pm$1     & 38$\pm$1       & 19.8$\pm$0.7       & 4.2$\pm$0.2 &  & \nodata \\   


 106252& obs           & 1200$\pm$9     & 746$\pm$9   & 271$\pm$2     & 92.8$\pm$0.9 & 30.6$\pm$0.3 & 30.6$\pm$0.6 & 18$\pm$1           & 16$\pm$9  & 1.4  & \nodata \\

& model                 & 1146$\pm$31    & 708$\pm$18  & 268$\pm$7     & 94$\pm$3   & 30.0$\pm$0.8 & 30.0$\pm$0.8   & 15.8$\pm$0.4       & 3.3$\pm$0.1& & \nodata \\   


 121504& obs           & 1002$\pm$7     & 631$\pm$8   & 225$\pm$1     & 81.3$\pm$0.8& 25$\pm$0.2  & 27$\pm$2       & 14$\pm$2           & 27$\pm$19 & 1.3    & \nodata \\

& model                 & 976$\pm$35     & 620$\pm$22  & 222$\pm$8     & 79$\pm$3   & 25$\pm$1     & 25$\pm$1      & 13.4$\pm$0.5       & 2.8$\pm$0.1  & & \nodata \\


 141937& obs            & 1393$\pm$10    & 872$\pm$11  & 311$\pm$2     & 110$\pm$1 & 34.9$\pm$0.3 & 35.1$\pm$0.5 & 18$\pm$2          & -3$\pm$12 & -0.6     & \nodata \\

& model                 & 1365$\pm$38    & 864$\pm$23  & 312$\pm$9     & 111$\pm$3  & 36$\pm$1     & 36$\pm$1       
& 18.7$\pm$0.5       & 4.0$\pm$0.1 & & \nodata \\   


 179949& obs           & 2943$\pm$21    & 1849$\pm$23 & 658$\pm$4     & 234$\pm$2 & 73.9$\pm$0.7 & 71$\pm$1    & 35$\pm$2           & -5$\pm$11 & -1.2   & 1$\pm$10 \\

& model                 & 2822$\pm$78    & 1809$\pm$49 & 641$\pm$18    & 229$\pm$6  & 73$\pm$2     & 73$\pm$2       & 38$\pm$1           & 8.1$\pm$0.2 & & \nodata \\  


 190228& obs           & 2068$\pm$15    & 1283$\pm$16 & 469$\pm$3     & 166$\pm$2 & 52.8$\pm$0.5 & 53.4$\pm$0.8 & 29$\pm$3          & 12$\pm$26 & 0.2   & \nodata \\

& model                 & 1987$\pm$66    & 1223$\pm$40 & 456$\pm$15    & 164$\pm$5  & 52$\pm$2     & 52$\pm$2      & 27.6$\pm$0.9       & 5.8$\pm$0.2  & & \nodata \\  


\tablenotetext{~}{NOTE. -- Photometry and 1-$\sigma$ internal uncertainties are in units of mJy. Calibration 
uncertainties are not included in the error estimates. IRS fluxes come from 
synthetic photometry from IRS low-resolution spectra. Fluxes at 1200 $\mu$m are from Carpenter et al. (2005).
(${\it model}$) is the expected stellar contribution from its Kurucz model; (${\it obs}$) is the photometric measurement. 
The signal-to-noise ratio of the excess, SNR$_{exc}$, is the photometric measurement minus the star's contribution from its Kurucz model divided by the global uncertainty. The global uncertainty is calculated adding in quadrature the internal and calibration uncertainties, the later taken to be 7\% for MIPS 70 $\mu$m.}
\enddata
\end{deluxetable}

\clearpage

%
%

\begin{deluxetable}{rrr}
\tablecaption{Relative Radial Velocities for HD~150706}
\label{}
\tablewidth{0pt}
\tablehead{
\colhead{JD}         & \colhead{RV}     & \colhead{Unc.}  \\
\colhead{-2450000}   & \colhead{(m/s)}  & \colhead{(m/s)}  }
\startdata
2514.728 &   -0.40 &    1.8  \\
2538.717 &   -3.59 &    1.9  \\
2713.091 &    4.20 &    1.7  \\
2806.008 &   -9.49 &    2.0  \\
2850.896 &   -5.08 &    1.8  \\
3179.924 &    0.00 &    1.7  \\
3427.128 &   23.83 &    2.2  \\
3842.061 &   20.89 &    2.2  \\
\enddata
\end{deluxetable}

\clearpage

\begin{figure}
\epsscale{0.9}
\plotone{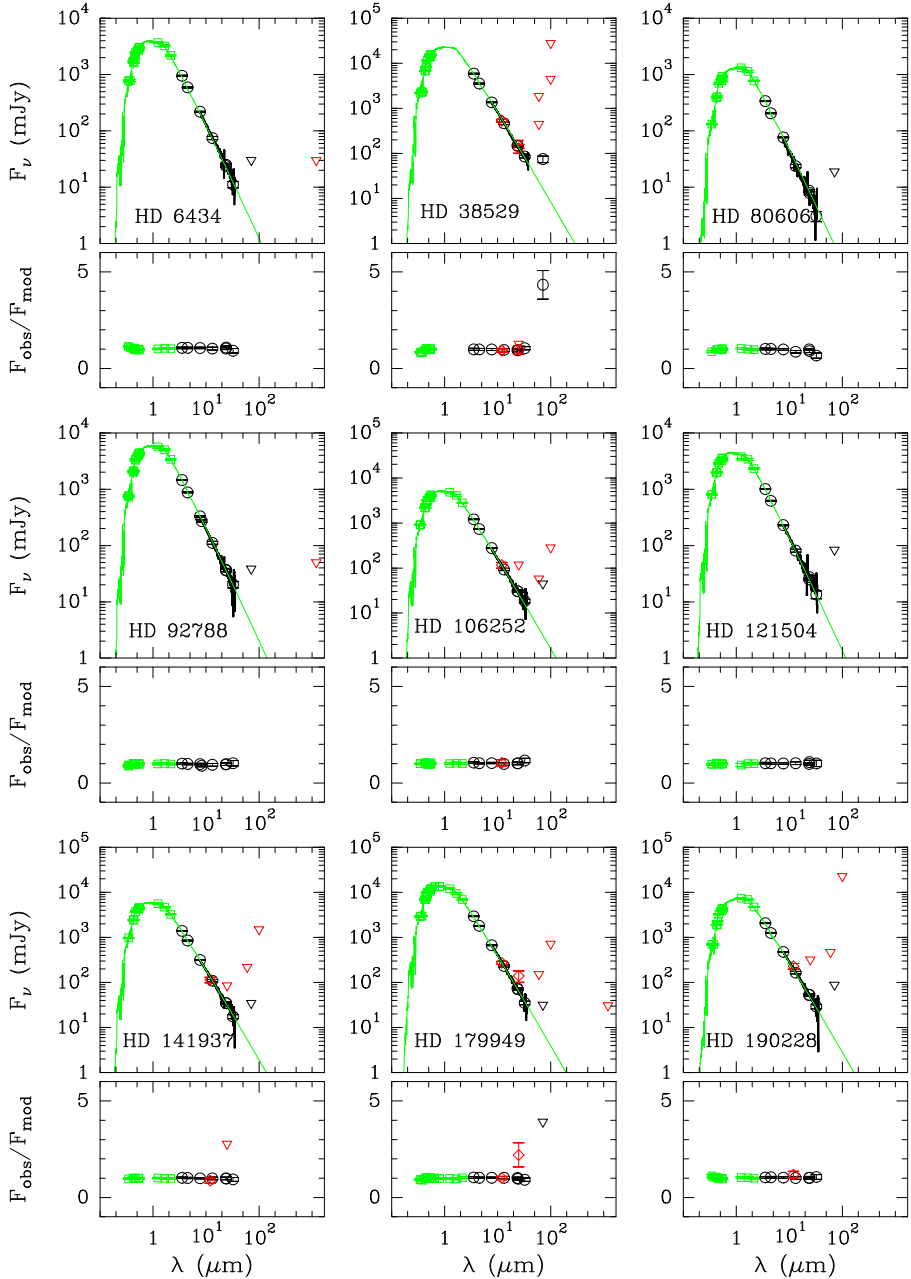}
\caption{Spectral energy distributions (SEDs) of the nine planet-bearing stars in the FEPS program. 
The green line is the Kurucz model. The black thicker line is the IRS low-resolution spectrum. The 
photometric points are identified as follows: green squares are ground based observations (including 
Tycho and 2MASS); black circles are $\it{Spitzer}$ observations (IRAC, MIPS and synthetic photometry 
from IRS); red diamonds are IRAS observations. In all cases, the error bars correspond to 1-$\sigma$ 
uncertainties.  Upper limits are represented by triangles and are given when F/$\Delta$F $<$ 3 and 
placed at F + 3$\times$$\Delta$F if F $>$ 0, or 3$\times$$\Delta$F if F  $<$ 0. Black triangles are 
upper limits for $\it{Spitzer}$ 70 $\mu$m, and red triangles for IRAS and 1.2 mm.}
\end{figure}

\clearpage

\begin{figure}
\epsscale{0.8}
\plotone{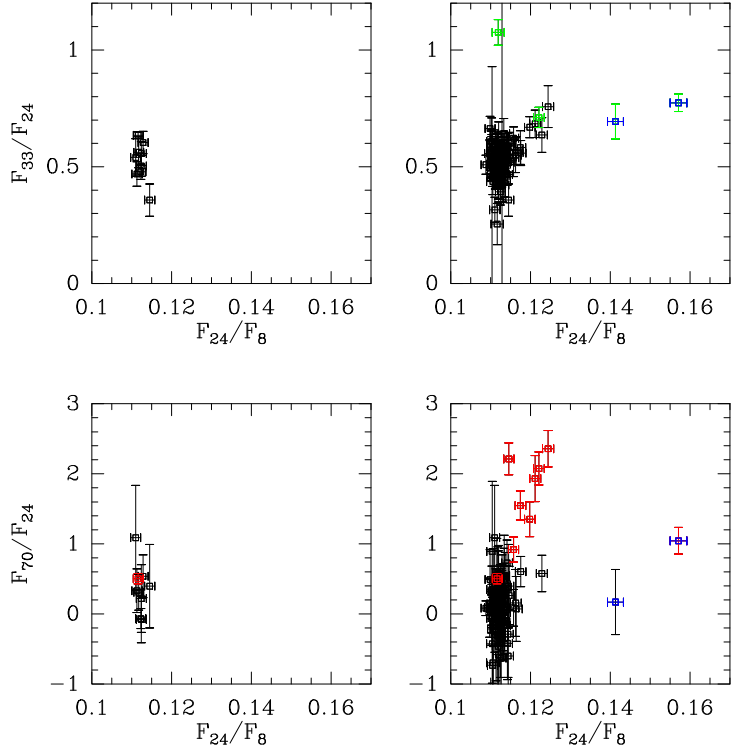}
\caption{Color-color diagrams of the 9 stars in the FEPS planet sample ($\it{left}$) and the 99 stars in the 
FEPS control sample ($\it{right}$). 
Stars that show 70 $\mu$m excess emission with a SNR in the excess $>$ 3 
are shown in $\it{red}$, and include one star in the planet sample (HD 38529) and nine stars in the control sample. 
Similarly, stars that show 24 $\mu$m and 33 $\mu$m excess emission with a SNR in the excess $>$ 3 are shown in 
$\it{blue}$ and $\it{green}$, respectively, and include only two stars (for 24 $\mu$m) and four stars 
(for 33 $\mu$m) in the control sample and none in the planet sample. [The outlier at (0.114, 0.357) in the 
upper left panel corresponds to HD 80606: its IRS spectrum
is very noisy beyond 30 microns, possibly due to a peak-up failure, 
making the 33 $\mu$m point unreliable].
}
\end{figure}

\clearpage

\begin{figure}
\epsscale{0.7}
\plotone{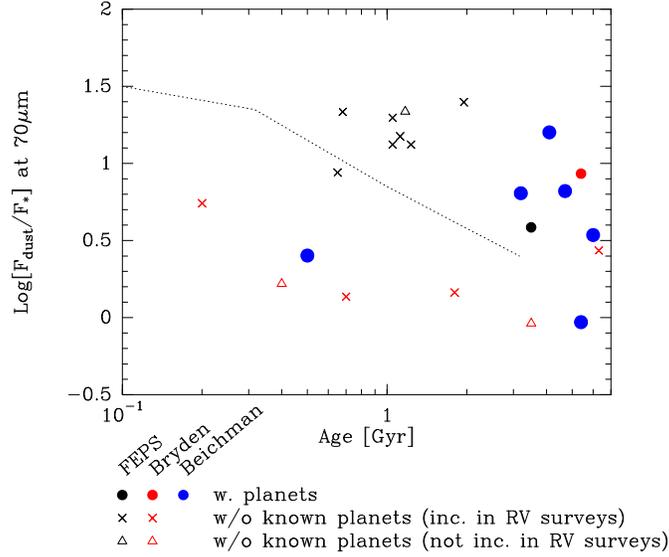}
\caption{Ratio of the excess flux to the photospheric flux for the stars with 70 $\mu$m excess emission 
and a SNR in the excess $>$ 3.
The shape of the symbol indicates the presence of a close-in 
planet. $\it{Circles}$ are stars with known radial velocity planets; $\it{crosses}$ are stars without known planets 
(but included in radial velocity surveys); and $\it{triangles}$: stars without known planets (not included in radial 
velocity surveys). The $\it{black}$ symbols correspond to stars in the FEPS survey, while 
$\it{blue}$ and $\it{red}$ correspond to stars in Beichman et al. (\citeyear{beic05}) and
Bryden et al. (\citeyear{bryd06}), respectively. (The latter includes stars at smaller distances
than those in the FEPS sample, so is sensitive to smaller excesses). For comparison, the $\it{dotted}$ $\it{line}$ shows 
F$_{dust}$/F$_{*}$ at 60 $\mu$m resulting from the collisional cascade of a planetesimal disk at 30--80 AU 
(Kenyon \& Bromley~\citeyear{keny05}). Because the collisional physics and the behavior of the debris following the 
collision are uncertain, the estimate for F$_{dust}$/F$_{*}$ at 60 $\mu$m could vary by more than a factor of 10, 
so the observations could be consistent with the model predictions.
} 
\end{figure}

\clearpage

\end{document}